\documentclass[twocolumn,preprintnumbers,amsmath,amssymb]{revtex4}

\usepackage{graphicx}
\usepackage{epstopdf}
\usepackage{dcolumn}
\usepackage{bm}
\usepackage{color}

\begin{document}
\def\P{\mathbf{P}}
\def\Q{\mathbf{Q}}

\title{Non-Equilibrium Thermodynamics of Self-Replicating Protocells}

\author{Harold Fellermann$^{1,2}$, Bernat Corominas-Murtra$^{3,4}$, Per Lyngs Hansen$^5$, John Hjort Ipsen$^5$,  Ricard Sol\'e$^{4,6}$ and Steen Rasmussen$^{2,6}$}
\affiliation{
$^1$ School of Computing Science, Newcastle University, Newcastle NE1 7RU, United Kingdom \\
$^2$Center for Fundamental Living Technology, University of Southern Denmark, Campusvej 55, 5230 Odense M, Denmark\\
$^3$ Section for Science of Complex Systems, Medical University of Vienna,  Spitalgasse 23, A-1090 Vienna, Austria\\
$^4$ICREA-Complex Systems Lab, Universitat Pompeu Fabra (GRIB), Dr Aiguader 80, 08003 Barcelona, Spain\\
$^5$Center for Membrane Biophysics, University of Southern Denmark, Campusvej 55, 5230 Odense M, Denmark\\
$^6$Santa Fe Institute, 1399 Hyde Park Road, Santa Fe NM 87501, USA}

\begin{abstract}
We provide a non-equilibrium thermodynamic description of the life-cycle of a droplet based, chemically feasible, system of protocells.
By coupling the protocells metabolic kinetics with its thermodynamics, we demonstrate how the system can be driven out of equilibrium to ensure protocell growth and replication.
This coupling allows us to derive the equations of evolution and to rigorously demonstrate how growth and replication life-cycle can be understood as a non-equilibrium thermodynamic cycle.
The process does not appeal to genetic information or inheritance, and is based only on non-equilibrium physics considerations.
Our non-equilibrium thermodynamic description of simple, yet realistic, processes of protocell growth and replication, represents an advance in our physical understanding of a central biological phenomenon both in connection to the origin of life and for modern biology.
\end{abstract}

\maketitle
{\em Introduction}.- Developing a physical understanding of the processes underlying biological phenomena is perhaps one of the greatest challenges for modern physics.
In this work we uncover the underpinning thermodynamics of cell growth and division in protocells composed of surfactant coated oil droplets in water.
Our objective is twofold: on the one hand, we try to disentangle the physical conditions for growth and division processes that are critical for all life.
On the other hand, an understanding of these phenomena may assist the ongoing work on assembling artificial cells in the laboratory, with a large number of potential technological applications.
Finally, we believe that a thermodynamic understanding of cell growth and division in one of its simplest implementations may facilitate our understanding of the more complex processes of modern cell division.

To bypass a discussion of the controversial topic about ``what is life'', we use an \emph{operational definition} of a living process as a physical entity~\cite{Ganti:2003, Sole:2007, Rasmussen:2008,KRMirazo:2013} that has the ability:
(i) to capture material resources and turn them into building blocks (grow and divide) by the use of external provided free energy (a metabolic machinery);
Hereby the system is driven out of equilibrium and should undergo a thermodynamic cycles every time it replicates \cite{Kauffman:2003};
(ii) to process, in part by controlling the metabolic processes under (i), and transmit (copy) inheritable information to progeny;
(iii) to keep its components together and distinguish itself from the environment (compartmentalization). The compartment contains the metabolic and the informational system;
(iv) to undergo Darwinian evolution through variation of the copied inheritable information and a successive selection of the better progeny.
Important advances have been made over the years regarding the thermodynamics of living processes~\cite{Morowitz:1968, Deamer:1997}.
Recently, additional advances on this topic has emerged, relating thermodynamics, information and the essential chemical reactions in living systems~\cite{Smith:2008a,Smith:2008b,Smith:2008c,England:2013}.

Here we present a non-equilibrium thermodynamic characterization by a system of protocells that are able to reproduce.
We device a physical self-replicating system with an energy transduction mechanism that converts chemical energy into mechanical energy that drives the aggregate division.
We address how the protocell stability is obtained in and out of equilibrium and how an instability is used to drive the protocellular self-replication.
In addition, the system is designed under realistic conditions to lead and reflect laboratory experiments in this area. Similar protocellular life-cycle systems have already been implemented in the lab both based on droplets and vesicles \cite{Maurer:2011, Caschera:2013}.
The system can be realized without any use of inheritable genetic information.
We thus have a system capable of reproducing (under external supply of chemical energy and matter) satisfying minimally conditions (i) and (iii) stated above.

\begin{figure*}
\includegraphics[width= 16.5cm]{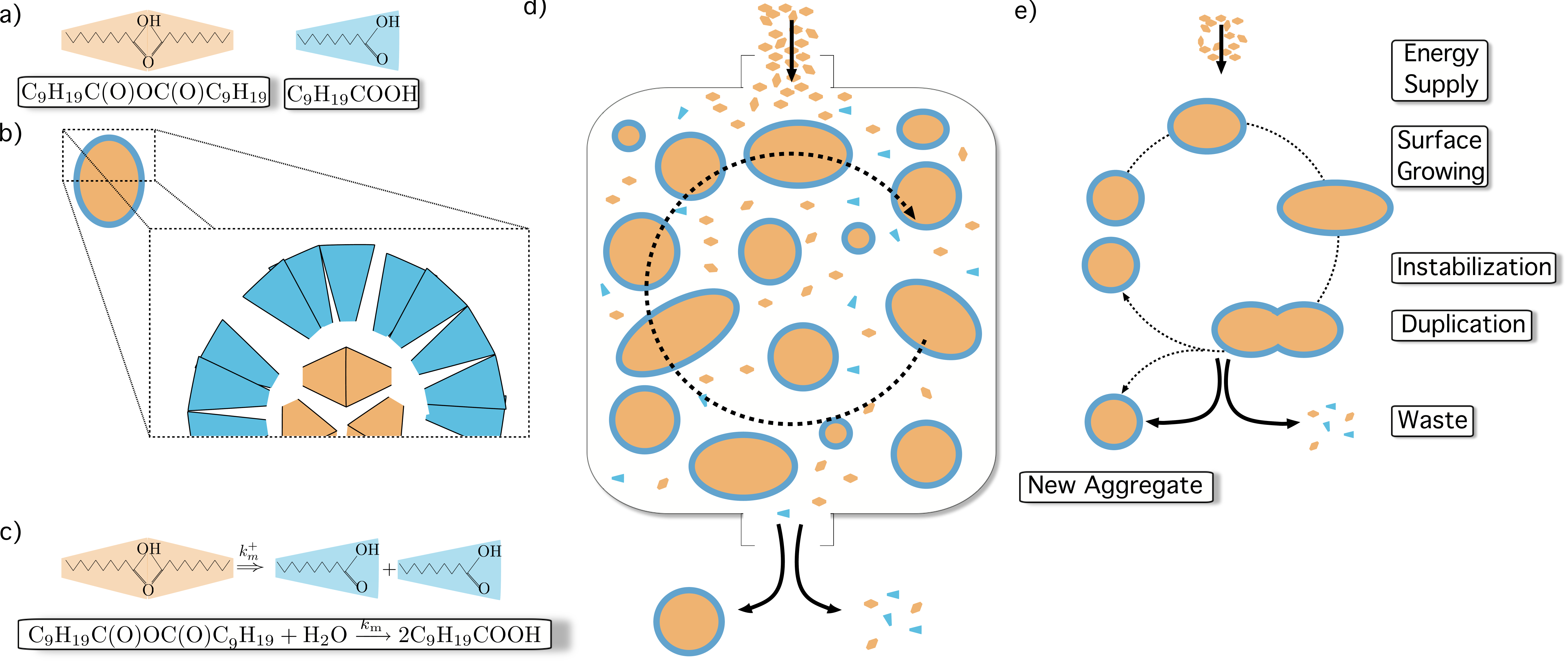}
\caption{
(a) Our system is composed of two types of molecules in an aqueous solution:
decanoic anhydrides (orange) and decanoic acid (blue).
(b) In aqueous solutions, the conic geometry of decanoic acid molecules (also referred to as surfactants or lipids) promotes aggregation of spherical clusters, due to the hydrophilic behaviour of the head and the hydrophobic behaviour of the tail.
The surfactant layer shields an encapsulated volume of decanoic anhydride, also referred to as precursors.
This creates a primitive compartment.
Notice that the cover of the surface might not be perfect.
(c) A molecule of anhydride spontaneously hydrolyzes two molecules of decanoic acid (blue) at a certain rate $k_\text m$.
This creates an extra availability of surface that deforms and eventually breaks the aggregates.
(d) Experimental setup: a reactor contains an emulsion with droplets composed of internal precursor molecules surrounded by lipids.
The reactor is constantly fed with precursors that incorporate into existing droplets.
The spontaneous metabolic reaction converts precursors into lipids, thus changing the droplets' surface to volume ratio until they become unstable and divide.
An outflow at the bottom removes newly created aggregates and waste, leaving the total aggregate density constant.
(e) A 'zoom' into a single replication cycle:
the chemical gradient of the input precursor molecules is the energetic inflow, whereas the newly created aggregate is the outcome of the cycle.
}
\label{fig_cycle_schematic}
\end{figure*}

As model system for simple replicating protocells we study oil-in-water emulsion compartments---ternary systems where a surfactant layer shields small hydrophobic volumes of oil molecules from their aqueous environment (i.e. Winsor type IV emulsion).
The amphipihlic surfactants lower the surface tension of the oil-water interface to a degree where suspended spherical droplets are thermodynamically stable due to the increase of mixing entropy in the system---see Figure \ref{fig_cycle_schematic}(a-b).
Surfactant molecules and, to a lesser extent, oil molecules will also be found in aqueous solution as this again increases the mixing entropy of the system.
We will refer to these compartments as \emph{oil droplets} or simply \emph{droplets}.
These droplets already mimic one of the fundamental properties of living beings, namely, property (iii), since it can act as a container for a metabolism and an information system and thus defines the boundary of the living system.
Emulsion compartments have been proposed previously as containers for experimental models of living systems \cite{Oparin:1953,Hanczyc:2007,Caschera:2013}.

We equip this model system with a simple metabolism by choosing an oil component that can be converted into surfactants.
Several such metabolisms have been proposed based on hydrolysis or photo-fragmentation of organic acid esters and anhydrides at the water-aggregate interface \cite{Bachmann:1992,DeClue:2009,Maurer:2011,Hanczyc:2011}.
Compared to lipid vesicles with \emph{encapsulated} metabolisms, the oil phase of emulsion compartments (as well as the exterior of a lipid membrane container) does not introduce a diffusive barrier to nutrient and waste fluxes toward and from the droplet interface.
Common to all these model metabolisms is that the hydrophobic compound serves as (in some of the above cases high-energy) nutrient that gets catabolized into an amphiphilic (in some of the above cases low energy) building block of the compartment plus a potential waste molecule.
For simplicity, but without loss of generality in our approach, we pick anhydride hydroplysis as the metabolic reaction as it avoids the introduction of a metabolic system with associated additional chemical species in the system.
In particular, we base our calculations on a system composed of decanoic acid surfactants C$_9$H$_{19}$COOH and decanoic anhydride precursors.
The metabolic reaction is then given as
\begin{equation}
	\mathrm{C}_9\mathrm{H}_{19}\mathrm{C(O)OC(O)C}_9\mathrm{H}_{19}  + \mathrm{H}_2\mathrm{O} \xrightarrow{k_\text m} 2 \mathrm{C}_9\mathrm{H}_{19}\mathrm{COOH},
\end{equation}
where the anhydride can be regarded as food molecule and the surfactant as building blocks.
The metabolic reaction continuously converts the hydrophobic volume of the droplets into new surfactants.
As the nutrient is hydrophobic but depends on the availability of water, we expect the reaction to take place at the compartment interface.
As a result of the changing surface to volume ratio, the aggregate will eventually become unstable and divide into smaller compartments, thereby accommodating for the newly produced surface molecules \cite{Mayer:2000,Fellermann:2007,Gao:2010}.

We now investigate a maintained aggregate feeding--division cycle and show that it has all the properties needed to be considered a non-equilibrium thermodynamic cycle, as it fulfils conditions (i) and (iii), while information replication (ii) and evolution (iv) are not part of this system.

{\em Thermodynamic landscape of the system}.- A state, $\sigma$, in this non-equilibrium thermodynamic system is described by five state variables
	$\sigma\equiv(L_\text d, P_\text d, L_\text b, P_\text b, n) \in \mathbb N^5$
which denote the number of surfactants $L$ and surfactant precursors $P$ arranged into $n$ oil droplets as well as in the bulk environment $L_\text b$ and $P_\text b$.
Alternatively, we write
\begin{equation}
	\sigma\equiv(L, P,n)_{(L_\text{tot},P_\text{tot})} = (L_\text d, P_\text d, L_\text b, P_\text b, n)
	\label{eq:State}
\end{equation}
to emphasize constant total numbers $L_\text{tot} = L_\text d+L_\text b$ and $P_\text{tot} = P_\text d+P_\text b$ (and we also simplify $L=L_\text d$, $P=P_\text d$).
We assume that the oil droplet can exchange particles, heat, and pressure with the environment (NPT ensemble) and that transfer of heat and pressure occur significantly faster than transfer of matter such that the system is instantaneously equilibrated with respect to temperature and pressure.

Aside from entropic contributions to be specified below, the change in (Gibbs) free energy associated with self-assembly of a droplet emulsion compartment from solution can be 
decomposed into the three components:
\begin{equation}
	G_\text{drop} = \Delta \mu_\text L L + \Delta \mu_{\text P} P + G_\text{geo}.
		\label{eq:Gdrop}
\end{equation}
where $\Delta\mu_\text L$ and $\Delta\mu_{\text P}$ are the changes in chemical potential when moving precursors and lipids from bulk into the aggregate, and $G_\text{geo}$ a geometric term expressing shape and surface contributions of the aggregate.
$\Delta\mu_\text L$ can be calculated from their partition coefficient---\emph{i.e.} the fraction of lipids found in bulk solution as opposed to the aggregates.
Bachman et al. estimate this value to be 14\% for surfactants with comparable solubility~\cite{Bachmann:1991}.
At $T=300 \text{K}$, this corresponds to $\Delta\mu_\text{L}=-4.53 \text{kJ/mol}$.
Since the anhydride has two hydrophobic chains, we set $\Delta\mu_\text{P}=2 \Delta\mu_\text{L} = -9.06 \text{kJ/mol}$, which in turn evaluates to a partition coefficient of 2.5\%.

To compute the geometric contribution to the energy, we observe that if the same principle of opposing forces~\cite{Israelachvili:1992} that dictates self-assembly of micelles will also drive assembly of droplet compartments, and if bending elastic contributions are ignored, the geometric contribution to the free energy reads (see supplementary information, SI):
\begin{equation}
	G_{\rm geo}=\gamma a+\frac{\beta}{a},
\end{equation}
where $\gamma$ is the surface tension, $\beta$ the compressibility coefficient, and $a$ the surface area of the compartment.
In the absence of precursor, the minimum of $G_\text{geo}$ where the opposing forces balance corresponds to $a_0= \sqrt{\beta/\gamma}$.
As a function of lipid molecules, $G_\text{Geo}$ can be expanded as (see SI):
\begin{equation}
	G_\text{geo}(L) \approx G_\text{geo}(L^\star) + \frac{a_0^2}{2} \sqrt{\frac{\gamma^3}\beta} (L-L^*)^2
	\label{eq_G_geo}
\end{equation}
where $L^\star$ is the number of molecules in the droplet at optimal packing.
In the presence of precursor, $\beta$ and $\gamma$ depend on the size of the core of the aggregate and thus the number of precursors, and a delicate competition between surfactant and precursor determines the coverage of either component in the droplet compartment.
Assuming a spherical oil core of $P_\text d$ precursor molecules---each with molecular volume $V_\text d=0.54 \text{nm}^3$---the optimal number of surfactant molecules with tail length $\ell=1.4 \text{nm}$ and effective head area $a_0=25 \text{\AA}^2$ is given as~\cite{Evans:1999}:
\begin{equation}
	L^\star(P_\text{d}) = \frac{4\pi}{a_0}\left( \left(\frac{3 V_\text{P}}{4\pi}P_\text{d}\right)^{1/3} + \ell \right)^2 .
\end{equation}
The surface tension parameter $\gamma$ can be evaluated from Langmuir trough measurements and equals $45.9 \text{mN/m}$ and $\beta$ equals $5.80 \times 10^{-45} \text{Nm}^3$.

Accounting properly for the degeneracy of states, and thus for translational and configurational entropies, the free energy of a \emph{system} in the state $\sigma=(L, P,n)_{(L_\text{tot},P_\text{tot})}$ becomes (see SI):
\begin{equation}
	\label{eq_G_system_Main}
	G(\sigma) = \mu^\circ_\text L L_\text{tot} + \mu^\circ_\text P P_\text{tot}
		+ n \; G_\text{drop}\left(\frac L n,\frac P n\right)-TS(\sigma),
\end{equation}
with the standard chemical potentials $\mu_L^\circ$ and $\mu_P^\circ$ of lipids and precursors, respectively, and
\begin{equation}
S(\sigma)= n k_\text B\log \frac{V}{V_\text P/e} \\
	+k_\text B \log\left[\left(\begin{matrix} L_\text{tot} \\ L\end{matrix}\right)\left(\begin{matrix} P_\text{tot} \\ P\end{matrix}\right)\right],
\end{equation}
being the translational and configurational entropy of the system at state $\sigma$,
where $V$ describes the system volume per droplet and $V_\text{P}$---the molecular volume of the precursor---has been chosen as typical volume unit.

Observing that emulsion droplets of typically $100 \text{nm}$ radius have a volume of $0.0040$ femtoliter, which---assuming a typical water-to-oil ratio of 10:1---gives a system volume of $0.044$ femtoliter per droplet.
This also implies that a milliliter of emulsion has an order of magnitude of $10^{13}$ oil droplets.
From the ratio of precursor to droplet volume, it follows that each droplet contains some $7,430,000$ molecules plus 2.5\% in bulk, totalling to about $P_\text{c}=7,620,000$ anhydride molecules.
With $L^\star(P_\text{d})$ and a partition coefficient of 14\%, this implies a total of $L_\text{c}=570,000$ surfactant molecules.
In other words, our emulsion consists of $9.1$ volume percent decanoic anhydride and $21 \text{mmol/l}$ decanoic acid.


Equation~\eqref{eq_G_system_Main} determines the thermodynamic landscape of the system for a given configuration $\left(L_\text{tot},P_\text{tot}\right)$.
Lipids and precursors enter and leave aggregates stochastically until their association and dissociation reaches detailed balance around an equilibrium $(L,P,n)_{L_\text{tot},P_\text{tot}}$.


So far we have described a system in equilibrium.
The metabolic reaction that converts supplied precursors into additional surfactants drives the systems out of equilibrium ---see Figure (\ref{fig_cycle_schematic}c).
\begin{align}
	{L_\text{tot}, P_\text{tot}} & \xrightarrow{k_\text m P}{L_\text{tot}\!+\!\nu, P_\text{tot}\!-\!1} .
\end{align}
Here, $\nu$ is the stoichiometric ratio of the metabolic reaction and equals $1$ for precursor esters and $2$ for anhydrides.
We assume that this covalent reaction is essentially irreversible.

If metabolic turnover operates significantly faster than the rearrangement of molecules as well as the fission and fusion of aggregates, the dynamics can be reasonably well approximated through a separation of time scales.
We thus consider the overall process as a series of equilibrium states 
under slowly moving boundary conditions.
However, if a separation of time scales is not possible we may use a Fokker-Planck type equation of evolution, as described in the SI.  A typical trajectory of the Fokker-Planck dynamics is shown in Fig 2(a). 

\begin{figure}[!ht]
\includegraphics[width=\columnwidth]{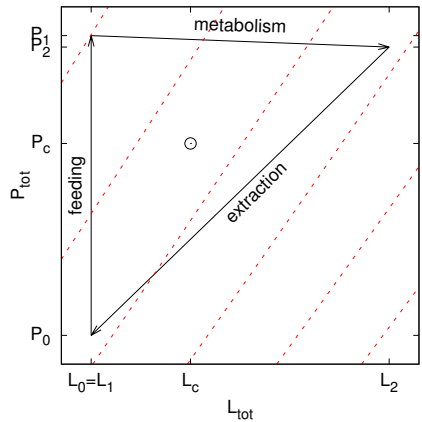}
\caption{
(Top) feeding-metabolism-extraction cycle of an emulsion in $\left(L_\text{tot},P_\text{tot}\right)$ space (solid black arrows) around the calibration point ($L_\text{c}, P_\text{c}$) where the system preferentially assembles into ten droplets.
Dashed red diagonal lines indicate configurations where the partitioning of matter into $n+1$ droplets is energetically equal to a partitioning into $n$ droplets, thus a single division or fusion event is expected each time an arrow crosses a diagonal.
Importantly, the number of metabolically induced divisions is one greater than the number of feeding induced fusions.
The newly created droplet resulting from the net process is extracted, in order to bring the system back to its initial condition.
(Bottom) Aggregate size given by the number of lipids ($L_\text d$) under periodic instantaneous feeding (red), compared to continuous feeding (blue).
Under continuous feeding, the aggregate undergoes a single division event without any fusion.
These numerical simulations confirm that a cycle is feasible with realistic parametrisation.
}
\label{fig_cycle_numerics}
\end{figure}

{\em Life cycles/Thermodynamic cycles}.- Our objective is to create stable conditions for the system to undergo successive thermodynamic cycles.
To this end, we provide a constant inflow of anhydrides to be used as building blocks of new droplets as well as precursors of surfactants through the metabolic reaction---see Figure (\ref{fig_cycle_schematic}d).
Let us assume a huge reservoir of $L_0,P_0$ total lipids and precursors, organized in its (local) equilibrium state in $n$ droplets.
We feed this reactor from the top with $\Delta P_\text{feed}$ precursors and water, such that it reaches the state $L_1,P_1=L_0,P_0\!+\!\Delta P_\text{feed}$.
Within the turnover time of the cycle, the reaction will convert a fraction $\Delta P$ of the precursors into $\Delta L=\nu\Delta P$ new surfactants.
This will drive the system into a new state consisting of $L_2,P_2=L_1\!+\!\Delta L, P_1\!-\!\Delta P$ lipids and precursors.
We compensate the inflow of precursors by an outflow that constantly removes from the system material proportional to one droplet
(containing $L_2/({n+1})$ lipid and $P_2/({n+1})$ precursor molecules) plus the proportional volume of aqueous solution.

If feeding and metabolism are tuned correctly, extraction of newly created droplets will compensate the inflow of precursors and the initial condition $L_0,P_0$ will be recovered---see Figure~\ref{fig_cycle_numerics} for the path through state space spanned by the actions of feeding, metabolism, and extraction.
We realize from the parametrization above that only a fraction $\phi$ about 4.4\% of the provided anhydrides are converted into additional surfactants.
Thus, in order to properly balance the metabolic reaction, feeding must proceed at an average rate $\left<k_\text{feed}\right> = \phi^{-1} k_\text{m} \left<P_\text{tot}\right>$.
Alternatively, the non-converted fraction of anhydride could be replaced by a non-reactive oil of comparable volume and hydrophobicity such as eicosane, $\mathrm{C}_{20}\mathrm{H}_{42}$, to decouple the cycle turnover time from the reaction speed.

The newly supplied precursors and surfactants change the equilibrium conditions, and the number of aggregates of the equilibrium state might change through aggregate division.
The change on the boundary conditions induced by the metabolic reaction can lead the system to the following scenario:
\begin{equation}
	G(\sigma_{n})>G(\sigma_{n+1}),
	\label{eq:ConditionDupl}
\end{equation}
where $\sigma_{n}$ and $\sigma_{n+1}$ are states of the system with the same amount of total molecules, $(L_2,P_2)$, but differing in the number of aggregates from $n$ to $n+1$.
If condition~\eqref{eq:ConditionDupl} is satisfied, duplication of aggregates is expected to occur.

At the microscopic level, the creation of new lipids will introduce a perturbation to the surface of existing aggregates, whose size will grow until, in some aggregate, the frustration due to the geometric term will drive it into an unstable state and, presumably, break in two small, more stable aggregates---see Figure \ref{fig_cycle_schematic}e.
Aggregates in this cycle are thus able to create new aggregates by division, and we say that aggregates exposed to these conditions \emph{self-replicate}.
Numerical simulations shown in Figure \ref{fig_cycle_numerics} indicate that cyclic processes of droplet division are expected under realistic parametrisation.

These boundary conditions maintain a stationary cycle, driven by the supply of precursors and extraction of waste, dissipated through the metabolic turnover into the eventual fusion of aggregates.
When a new aggregate is created, we say that a \emph{life cycle} has been completed.
We will refer to such cycle as $\omega$.
For the sake of discussion, we consider the initial point to be an equilibrium state with $n$ aggregates.
After feeding and metabolic turnover the system rearranges into a new equilibrium state containing $n+1$ aggregates, one of which subsequently being expelled by the boundary conditions to reconstitute the initial condition.
We recognize the newly created (and expelled) aggregate as the \emph{outcome} of the life cycle.
Since $\omega$ is a closed path over a potential function, we have that
\begin{equation}
	\Delta G_\text{feeding} + \Delta G_\text{metabolism} + \Delta G_\text{extraction} = 0,
\end{equation}
from which we can derive the energy change associated with the replication process as
\begin{align}
	\Delta G_\text{rep} &= -\Delta G_\text{feeding} - \Delta G_\text{extraction} \nonumber \\
	&= G(L_2,P_2) - G(L_1,P_1) \nonumber \\
	&= \Delta G_\text{metabolism} = \Delta G_{P \rightarrow \nu L} \Delta P \label{eq_work},
\end{align}
which depends on the exact change in chemical potential $\Delta G_{P \rightarrow \nu L} = \nu\mu^\circ_\text L - \mu^\circ_\text P$ associated with the metabolic reaction.
For typical anhydride hydrolyses reactions, $\Delta G_{P \rightarrow \nu L}$ is on the order of magnitude of $-10 \text{kJ/mol}$, this resolves to an order of magnitude of $-0.1 \text{J}$ per millilitre of produced (extracted) emulsion.

Since anhydride hydrolysis is an exergonic reaction that proceeds spontaneously, the energy change of equation~\eqref{eq_work} associated with replication should not be conceptualized as work---of which we would only speak if the energy of the downhill reaction would be harvested to drive an endergonic reaction against it's natural direction.
Thus, in this situation we encounter a self-constructing system that spontaneously creates order, or ``constraints'', by self-assembly from a higher energy state~\cite{Kauffman:2003}.
Alternatively, had we used a protocellular metabolism that requires external pumping of free energy, e.g. the photo-fragmentation reaction \cite{DeClue:2009,Maurer:2011}, the necessary photo-energy per produced fatty acid would be $\sim 2.1 \text{eV}$, which corresponds to $\sim 200 \text{kJ/mol}$, and the life-cycle could be conceptualised as a work-cycle. Using the same assumptions as above, this resolves to an order of magnitude of $20 \text{J}$ per millilitre of produced (extracted) emulsion.

{\em Discussion}.- The full thermodynamic characterisation of a life cycle represents a further step towards the understanding life as a physical phenomenon.
We have shown that, under realistic assumptions, certain chemical systems are expected to display the onset of {\em biological behaviour}.
Indeed: we have shown the feasibility and physical consistency of an oil droplet container system to be able to grow and reproduce---i.e., to perform a thermodynamic cycle---in accordance with the laws of thermodynamics.
The growing process represents the ability of the system to take material from the outside to be used as building blocks.
The replication, driven by energy unbalances between geometrical configurations, implies the possibility of creating a {\em population} of aggregates which may grow and expand as long as the physical conditions are favourable.
Our results point to the conception of life as an expected emerging phenomena from non-living chemical substances under special physical conditions of matter gradients and appropriate energy flows.

{\em Acknowledgemnts}.- Anders Andersen is acknowledged for helpful discussion.
HF and SR received financial support from the Danish National Research Foundation and the European Commission sponsored projects MATCHIT and MICREagents. BC-M acknowledges the hospitality of the Center for Fundamental Living Technology/University of southern Denmark and the financial support from the Austrian Science Fund FWF under KPP23378FW and the Marcelino Bot\'in Foundation.


\newpage
\appendix
\section{}

\subsection{The geometric contribution to the free energy}

In general, the free energy of an aggregate is a three-term function depending on its area and the set of parameters accounting for the membrane properties of the system:
\[
G_{\rm geo}=\gamma a+\frac{\beta}{a}+\kappa\oint_{a}(H-H_0)^2da,
\]
where $\gamma$ is the surface tension, $\beta$ the compressibility coefficient, and $\kappa$ the elastic bending modulus.
We consider that $\gamma,\beta\gg\kappa$ thus we can neglect the contribution of the Helfrich Hamiltonian $\kappa\oint_a ...$, resulting in a model of opposing forces.
The minimum of $G_{\rm geo}$ is found at:
\[
\frac{\partial}{\partial a}G_{\rm geo}=0;\;\;\rightarrow\;\;a_0=\sqrt{\frac{\beta}{\gamma}}.
\]
It turns out that, if we have the area of the head of the surfactant molecules, we can compute the ideal coverage number of surfactants $L^*$ from the ideal area $a_0$ as:
\[
L^{\star}=\sqrt{\frac{\beta}{\gamma}}\frac{1}{a_0},
\]
where $a_0$ is the effective head area of the surfactant molecules.
Now we interpret the role of parameters $\beta$ and $\gamma$ as depending on the crowding of the core of the aggregate, therefore, $L^{\star}\equiv L^{\star}(P)$.

%

Let us assume that we are close to the equilibrium.
If this is the case, we can compute increases of free energy through a Taylor approach:
\begin{align*}
	G(a) &= G(a_0) + \frac{1}{2} \left.\frac{d^2G}{da^2}\right|_{a_0}(a-a_0)^2 \\
	\Delta G(L) &= \frac{a_0^2}{2} \sqrt{\frac{\gamma^3}\beta} (L-L^*)^2 \\
\end{align*}

The compressibility coefficient $\beta$ can be derived from the second virial coefficient:
assuming that a virial expansion is appropriate for the 2D surfactant layer that covers the aggregate, the equation of state for the 2D pressure P will contain terms of the form
\[
\frac{P\cdot a}{L k_\text{B}T} = 1 + B_2 L/a + B_3 (L/a)^2 + \ldots
\]
The first term is an ideal gas term and the third term is ignored here (or could enter in a re-definition of $B_2$).
Thus we focus on the second $B_2$ term (the second-virial correction) in which the virial coefficient $B_2$ is expected to be of order $a_0^2$.
This statement can be made more precise via classical stat. mech. modeling of the surfactant layer or as we do now:

Assuming the above equation of state, the compression term in the Gibbs energy will contain terms such as
\[
G_\text{compression} = -\int (P(a)da) 
                        \longrightarrow G_\text{ideal} + B_2k_\text{B}T (L^2/a) + \ldots,
\]
and the non-ideal term $B_2 k_\text{B}T (L^2/a)$ will contribute to the compressibility $K$ as follows:
\[
K^{-1} = - a\frac{\partial P}{\partial a} = - a_0 \frac{\partial P}{\partial a} 
                      \longrightarrow K^{-1}_\text{ideal} + 2 B_2 k_\text{B}T (L/a_0)^2
\]
We may then \emph{define} $B_2 k_\text{B}T$ operationally to mean,
\[
2 B_2 k_\text{B}T  =  a_0 \frac{\partial P}{\partial a}  (a/L)^2 
                       \simeq L^\star a_0^3  \frac{\partial P}{\partial a} 
\]
and thus $\beta$ resolves to:
\begin{equation}
	\beta = \frac{a_0^3}{2{L^\star}^2} \frac{\partial P}{\partial a}
	\label{eq:beta}
\end{equation}

To compute $L^*$, let $V_P$ be the volume of a precursor molecule and $a_0, \ell$ be the head area and length of a lipid molecule.
Then $V = V_P P$ is the volume of a sphere consisting of $P$ precursor molecules and
\[
	r = \left(\frac{3}{4\pi} V_P  P\right)^{1/3}
\]
its radius.
The radius of the entire aggregate (precursor core plus lipid layer) is $r+l$.
Consequently, the surface area of the aggregate is
\[
	a = 4\pi (r+\ell)^2 = 4\pi \left( \left(\frac{3}{4\pi} V_P  P\right)^{1/3} + \ell\right)^2 .
\]
The ideal number of lipids required to cover the surface area $a$ is given by
\[
	L^\star = a/a_0 = \frac{4\pi}{a_0}\left( \left(\frac{3}{4\pi} V_P  P\right)^{1/3} + \ell\right)^2 .
\]

\begin{widetext}

\subsection{Equations of Evolution}

{\bf Transitions} Below we detail all the transitions occurring in our system.
A state $\sigma$ is completely described by five variables, $\sigma\equiv(L, P,n)_{(L_\text{tot},P_\text{tot})}$ as described in equation (\ref{eq:State}) of the main text.
The free energy of the system in the state $\sigma$, $G(\sigma)$ is defined in the equation (\ref{eq_G_system_Main}) of the main text.
Here we adopted the notation $G(L,P,n)\equiv G(\sigma)$, since we have the need to explicitly express the changes on the variables.
\begin{eqnarray}
	(L, P,n)_{(L_\text{tot},P_\text{tot})} &\xrightarrow{k_\text n^+ n} &(L, P,n+1)_{(L_\text{tot},P_\text{tot})} \nonumber\\
	(L, P,n)_{(L_\text{tot},P_\text{tot})} &\xrightarrow{k_\text n^- n} &(L, P,n-1)_{(L_\text{tot},P_\text{tot})};
		 k_\text n^ - =k_\text n^ + e^{(G(L,P,n-1)-G(L,P,n))/k_BT} \nonumber\\
	(L, P,n)_{(L_\text{tot},P_\text{tot})} &\xrightarrow{k_\text L^+ L_\text b} &(L+1, P,n)_{(L_\text{tot},P_\text{tot})} \nonumber\\
	(L, P,n)_{(L_\text{tot},P_\text{tot})} &\xrightarrow{k_\text L^- L} &(L-1, P,n)_{(L_\text{tot},P_\text{tot})};
		 k_\text L^ - =k_\text L^ + e^{(G(L-1,P,n)-G(L,P,n))/k_BT} \nonumber\\
	(L, P,n)_{(L_\text{tot},P_\text{tot})} &\xrightarrow{k_{\text P}^+ P_\text b} &(L, P+1,n)_{(L_\text{tot},P_\text{tot})}\nonumber \\
	(L, P,n)_{(L_\text{tot},P_\text{tot})} &\xrightarrow{k_{\text P}^- P} &(L, P-1,n)_{(L_\text{tot},P_\text{tot})};
		k_{\text P}^-=k_{\text P}^+e^{(G(L,P-1,n)-G(L,P,n))/kT}\nonumber\\
	(L, P,n)_{(L_\text{tot},P_\text{tot})} &\xrightarrow{k_\text m P} &(L+1, P-1,n)_{(L_\text{tot}+1,P_\text{tot}-1)} \nonumber
\end{eqnarray}

{\bf The master equation} The above described chemical relations can be inserted into a master equation.
For the sake of readability,  we rewrite $x\equiv P$, $x_0\equiv P_\text{tot}$, $y\equiv L$, $y_0\equiv L_\text{tot}$, $z\equiv n$,  $\sigma=(x,y,z,x_0, y_0)$.
In addition, $\mathbb{P}(\sigma)=\mathbb{P}(x,y,z,x_0, y_0)$, and we will use the explicit form to avoid any confusion.
We then define $\gamma(x,y,z)\equiv G(\sigma)/k_BT$.
The continuous version of the Master equation obtained from the above transitions between states equation reads:
\begin{eqnarray}
\frac{\partial}{\partial t}\mathbb{P}(x,y,z,x_0, y_0)&=&k_n^+\left\{(z+\delta z)\mathbb{P}(x,y,z+\delta z,x_0,y_0)-z\mathbb{P}(x,y,z,x_0,y_0) +\right.\nonumber\\
&&+\left. (z+\delta z)e^{\gamma(x,y,z+\delta z)-\gamma(x,y,z,x)}\mathbb{P}(x,y,z+\delta z)-ze^{\gamma(x,y,z)-
\gamma(x,y,z-\delta z)}\mathbb{P}(x,y,z,x_0,y_0)  \right\}\nonumber\\
&&+k_\text m\left\{(x+\delta x)\mathbb{P}(x+\delta x, y-\delta y, z, x_0-\delta x_0, y_0+\delta y_0 )- x\mathbb{P}(x,y,z, x_0, y_0)\right\}\nonumber\\
&&+k_{\text P}^+\left\{(x_0-x+\delta x)\mathbb{P}(x-\delta x,y,z,x_0,y_0)-(x_0-x)\mathbb{P}(x,y,z,x_0,y_0)+\right.\nonumber\\
&&\left.+(x+\delta x)e^{\gamma(x,y,z)-\gamma(x+\delta x,y,z)}\mathbb{P}(x+\delta x,y,x_0,y_0)-xe^{\gamma(x-\delta x, y,z)
-\gamma(x,y,z)}\mathbb{P}(x,y,z,x_0,y_0)\right\}\nonumber\\
&&+k_{\text L}^+\left\{(y_0-y+\delta y)\mathbb{P}(x,y-\delta y,z,x_0,y_0)-(y_0-y)\mathbb{P}(x,y,z,x_0,y_0)\right.\nonumber\\
&&\left.+(y+\delta y)e^{\gamma(x,y,z)-\gamma(x,y+\delta y,z)}\mathbb{P}(x ,y+\delta y,z,x_0,y_0)-ye^{\gamma(x, y-\delta y)-\gamma(x,y)}\mathbb{P}(x,y,x_0,y_0)\right\}.\nonumber
\label{continuousMasterEq}
\end{eqnarray}

{\bf Expansion of the Master equation}.- We expand the above master equation up to second order to obtain a differential operator accounting for the evolution of the system.
We divide the expansion in the three parts corresponding to the coordinates $x,y,z$.
$\\$
$\\$
$\\$
\noindent
{\em Precursors Term}.-
Let us rewrite the $x$-coordinate of the master equation (we do not explicitly write the dependence of the functionals on the other variables, for the sake of readability):
\begin{eqnarray}
\frac{\partial}{\partial t}\mathbb{P}(x)&=&k_{\text P}^+\left\{(x_0-x+\delta x)\mathbb{P}(x-\delta x)-(x_0-x)\mathbb{P}(x)+\right.\nonumber\\
&&\left.+(x+\delta x)e^{\gamma(x)-\gamma(x+\delta x)}\mathbb{P}(x+\delta x)-xe^{\gamma(x-\delta x)-\gamma(x)}\mathbb{P}(x)\right\}\nonumber
\end{eqnarray}
We can expand the exponential up to first order, namely:
\begin{eqnarray}
e^{\gamma(x)-\gamma(x+\delta x)}=1-\frac{\partial \gamma(x)}{\partial x}\delta x+{\cal O}(\delta x^2),\nonumber
\end{eqnarray}
to obtain
\begin{eqnarray}
\frac{\partial}{\partial t}\mathbb{P}(x)&=&k_{\text P}^+\left\{(x_0-x+\delta x)\mathbb{P}(x-\delta x)-(x_0-x)\mathbb{P}(x)+\right.\nonumber\\
&&\left.+(x+\delta x)\left[1-\frac{\partial \gamma(x,y,z)}{\partial x}\delta x\right]\mathbb{P}(x+\delta x)-x\left[1-\frac{\partial \gamma(x,y,z)}{\partial x}\delta x\right]\mathbb{P}(x)\right\}.\nonumber
\end{eqnarray}
Rearraging terms, we have that:
\begin{eqnarray}
\frac{1}{k_{\text P}^+}\frac{\partial}{\partial t}\mathbb{P}(x)&=&x_0(\mathbb{P}(x-\delta x)-\mathbb{P}(x))+(x+\delta x)\mathbb{P}(x+\delta x)-
(x-\delta x)\mathbb{P}(x -\delta x)+\nonumber\\
&&+(x+\delta x)\frac{\partial \gamma(x)}{\partial x}\delta x\mathbb{P}(x+\delta x)-x\frac{\partial \gamma(x)}{\partial x}\delta x\mathbb{P}(x)\nonumber
\end{eqnarray}
Now we expand the differences up to the second order:
\[
x_0(\mathbb{P}(x-\delta x)-\mathbb{P}(x))=x_0\left[-\frac{\partial}{\partial x}\delta x+\frac{\partial^2}{\partial x^2}\delta x^2+{\cal O}(\delta x^3)\right]\mathbb{P}(x).
\]
The second term is more tricky.
Indeed, whereas above the standard definition of derivative,
\[
\frac{f(x+\delta x)-f(x)}{\delta x}=f'(x)+{\cal O}(\delta x),
\]
works, now we have a \emph{midpoint derivative}, i.e.:
\[
\frac{f(x+\delta x)-f(x-\delta x)}{2\delta x}=f'(x)+{\cal O}(\delta x^2).
\]
The crucial observation is that we already have a second order correction in such a first derivative, thus we only have to perform one derivative to have an approach up to the second order of the difference.
Therefore:
\[
(x+\delta x)\mathbb{P}(x+\delta x)-(x-\delta x)\mathbb{P}(x -\delta x)=2x\left[\frac{\partial }{\partial x}\delta x+{\cal O}(\delta x^3)\right]\mathbb{P}(x).
\]
And the remaining term is expanded as follows:
\begin{eqnarray}
(x+\delta x)\frac{\partial \gamma(x)}{\partial x}\delta x\mathbb{P}(x+\delta x)+x\frac{\partial \gamma(x)}{\partial x}\delta x\mathbb{P}(x)=\left[x\frac{\partial \gamma(x)}{\partial x}\delta x\frac{\partial }{\partial x}\delta x+{\cal O}(\delta x^3)\right]\mathbb{P}(x)\nonumber.
\end{eqnarray}
Therefore, we have that
\[
\frac{1}{k_{\text P}^+}\frac{\partial}{\partial t}\mathbb{P}(x)=\left[\left(2x-x_0-x\frac{\partial \gamma(x)}{\partial x}\right)\frac{\partial }{\partial x}+
\frac{x_0}{2}\frac{\partial^2}{\partial x^2}\right]\mathbb{P}(x).
\]
$\\$
$\\$
$\\$
\noindent
{\em Lipids Term}.- It is easy to see that the $y$ coordinate behaves exactly as the $x$-coordinate up to a constant.
Thus,
\[
\frac{1}{k_{\text L}^+}\frac{\partial}{\partial t}\mathbb{P}(y)=\left[\left(2y-y_0-y\frac{\partial \gamma(y)}{\partial y}\right)\frac{\partial }{\partial y}+
\frac{y_0}{2}\frac{\partial^2}{\partial y^2}\right]\mathbb{P}(y).
\]
$\\$
$\\$
$\\$
\noindent
{\em Number of aggregates term}.-
$z$ coordinate evolves differently from the $x,y$ ones.
The evolution of this coordinate is driven by:
\[
\frac{1}{k^+_n}\frac{\partial}{\partial t}\mathbb{P}(z)=(z-\delta z)\mathbb{P}(z-\delta z)-z\mathbb{P}(z)+(z+\delta z)e^{\gamma(z+\delta z)-\gamma(z)}-
ze^{\gamma(z)-\gamma(z-\delta z)}\mathbb{P}(z).
\]
The first term is expanded as:
\[
(z-\delta z)\mathbb{P}(z-\delta z)-z\mathbb{P}(z)=\left[-z\frac{\partial }{\partial z}\delta z+\frac{z}{2}\frac{\partial^2 }{\partial z^2}\delta z^2\right]\mathbb{P}(z),
\]
and the second one:
\[
(z+\delta z)e^{\gamma(z+\delta z)-\gamma(z)}-ze^{\gamma(z)-\gamma(z-\delta z)}\mathbb{P}(z)=\left(1+\frac{\partial \gamma(z)}{\partial z}\delta z\right)\left[
z\frac{\partial}{\partial z}\delta z+\frac{z}{2}\frac{\partial^2}{\partial z^2}\delta z^2\right]\mathbb{P}(z).
\]
If we keep terms up to second order, we reach the following Fokker-Planck like equation:
\[
\frac{1}{k^+_n}\frac{\partial}{\partial t}\mathbb{P}(z)=\left[z\frac{\partial \gamma(z)}{\partial z}\frac{\partial}{\partial z}+z\frac{\partial^2}{\partial z^2}\right]\mathbb{P}(z).
\]
$\\$
$\\$
$\\$
\noindent
{\em Metabolic Term}.- 
The contribution of the metabolic turnover to the equation of evolution represented by the master equation is given by the following difference:
\[
k_\text m\left\{(x+\delta x)\mathbb{P}(x+\delta x, y-\delta y, z, x_0-\delta x_0, y_0+\delta y_0 )- x\mathbb{P}(x,y,z, x_0, y_0)\right\}.
\]
We observe that in this case we explicitly take into account the total number of lipids and precursors, $x_0, y_0$.
To obtain a differential operator accounting for the contribution of the metabolic part to the main equation of evolution, we observe that the difference above reported must be approached by a \emph{directional derivative}, $\nabla_{\vec{e}}$.
If $f:\mathbb{R}^n\to \mathbb{R}$ and $\vec{e}\in \mathbb{R}^n$, the directional derivative of $f$ along the \emph{direction} given by the vector $\vec{e}$ is given by:
\[
\vec{\nabla}_{\vec{e}}f=\frac{f(\vec{x}+\delta_{\vec{e}} x)-f(\vec{x})}{\delta x}+{\cal O}(\delta).
\]
We observe that $\vec{\nabla}_{\vec{e}}f=\vec{\nabla} f\cdot \vec{e}$.
in our case, $n=5$ and we perform the following change of variables for the sake of simplicity: $x=x_1, y=x_2, z=x_3, x_0=x_4, y_0=x_5$.
Therefore, the expansion will be performed over the following difference:
\[
(x_1+\delta x_1)\mathbb{P}(x_1+\delta x_1, x_2-\delta x_2, x_3, x_4-\delta x_4, x_5+\delta x_5)- x_1\mathbb{P}(x_1,x_2,x_3, x_4, x_5).
\]
If $\vec{x}\equiv(x_1,x_2,x_3, x_4, x_5)$, one can rewrite the above expression in a compressed way, namely:
\[
(x_1+\delta x_1)\mathbb{P}(\vec{x}+\delta_{\vec{e}} x)-x_1\mathbb{P}(\vec{x})=x_1\vec{\nabla}_{\vec{e}}\mathbb{P}(\vec{x})+{\cal O}(\delta),
\]
being
\[
\vec{e}=(1,-1,0,-1,1).
\]
From the observation that  $\vec{\nabla}_{\vec{e}}f=\vec{\nabla} f\cdot \vec{e}$, higher order terms are easily computed, leading to a second order term of the expansion like:
\[
\sum_{i,j\leq 5} x_1\frac{\partial^2}{\partial x_i\partial x_j}e_ie_j.
\]
$e_i$ and $e_j$ are the $i$th and $j$th components of the vector $\vec{e}$, respectively.
Collecting the first and second order approximations we get the final shape of the differential operator accounting for the role of the metabolism in our system, ${\cal L}$, namely:
\begin{equation}
{\cal L}= k_\text m\left(x_1\vec{\nabla}_{\vec{e}}+\frac{1}{2}\sum_{i,j\leq 5} x_1\frac{\partial^2}{\partial x_i\partial x_j}e_ie_j\right).\nonumber
\end{equation}

{\bf The Equation of Evolution for the System} Collecting all the above derivations, one has that the equation of evolution of the system is given by:
\[
\frac{\partial}{\partial t}\mathbb{P}(\sigma)=\left(\vec{\mathbf{u}}\cdot\vec{\nabla}+\vec{\mathbf{v}}\cdot\vec{\partial^2}+{\cal L}\right)\mathbb{P}(\sigma).
\label{eq:GeneralEq}
\]
where:
\begin{eqnarray}
\vec{\mathbf{u}}&=&\left(k^+_\text P \left[2x-x_0-x\frac{\partial}{\partial x}\gamma(x,y,z)\right],k^+_\text L\left[2y-y_0-y\frac{\partial}{\partial y}\gamma(x,y,z)\right], 
k^+_n\left[z\frac{\partial}{\partial z}\gamma(x,y,z)\right]\right)\nonumber\\
\vec{\mathbf{v}}&=&\left(k^+_\text P\frac{x_0}{2},k^+_\text L\frac{y_0}{2},k^+_n z\right)\nonumber\\
\vec{\nabla}&=&\left(\frac{\partial}{\partial x},\frac{\partial}{\partial y},\frac{\partial}{\partial z}\right)\nonumber\\
\vec{\partial ^2}&=&\left(\frac{\partial^2}{\partial x^2},\frac{\partial^2}{\partial y^2},\frac{\partial^2}{\partial z^2}\right)\nonumber
\end{eqnarray}
and, making the temporary change of notation $x=x_1,y=x_2,z=x_3,x_0=x_4,y_0=x_5$, for the sake of simplicity, ${\cal L}$ reads:
\begin{eqnarray}
{\cal L}= x_1\vec{\nabla}_{\vec{e}}+\frac{1}{2}\sum_{i,j\leq 5} x_1\frac{\partial^2}{\partial x_i\partial x_j}e_ie_j,\nonumber
\label{LM}
\end{eqnarray}
where $\vec{\nabla}_{\vec{e}}$ is the \emph{directional derivative} of the scalar field defined by $\mathbb{P}(x_1,x_2,x_3,x_4,x_5)$ along the vector $\vec{e}=(1,-1,0,-1,1)$.
$e_i, e_j$  are the $i$th and $j$th components of the vector $\vec{e}$.
We observe that the $5$ variables of the system are coupled and no reduction of dimension can be performed without making assumptions.
Since the time scales between metabolism and association-disassociation processes are such that:
\[
	k_\text m\ll k_{\text P}^+,k_{\text L}^+, k^+_\text n
\]
we can consider that the equilibration of the association-disassociation processes is faster enough to consider that metabolism \emph{always acts} over equilibrated aggregates.
This mathematically implies that we can neglect the contribution of ${\cal L}$, leaving the variation of the total number of molecules -due to the irreversible process of metabolism- as the initial conditions of the following Fokker-Planck like equation:
\[
\frac{\partial}{\partial t}\mathbb{P}(\sigma)=\left(\vec{\mathbf{u}}\cdot\vec{\nabla}+\vec{\mathbf{v}}\cdot\vec{\partial^2}\right)\mathbb{P}(\sigma).
\]

\end{widetext}

\subsection{Calibration and numerical solution}
Decanoic anhydride has a molecular volume of $V_\text{P}=0.54 \text{nm}^3$, whilst the hydrocarbon chain of decanoic acid has a length of $\ell=1.4 \text{nm}$.
We assume an effective head area $a_0=25 \text{\AA}^2$ \cite{Evans:1999}.
With these measurements, we can calculate the ideal number $L^\star$ of surfactants required to cover a spherical oil droplet containing $P_\text d$ anhydride molecules assuming perfect packing:
\[
	L^\star(P_\text{d}) = \frac{4\pi}{a_0}\left( \left(\frac{3 V_\text{P}}{4\pi}P_\text{d}\right)^{1/3} + \ell \right)^2 .
\]
The measurements imply a surfactant packing parameter of $\frac{V_\text{L}}{a_0\ell}=0.77$ and a mean aggregation number of decanoic acid micelles $L^\star(0)$ resolving to 98 molecules, which agrees well with reported values \cite{Anianson:1976}

As described in the main text, we penalize aggregates that deviate from the perfect covering with a harmonic expansion around the minimum (zero) energy value:
\[
	\Delta G(L) = G(L^*) + \frac{a_0^2}{2} \sqrt{\frac{\gamma^3}\beta} (L-L^*)^2
\]

To determine $G_\text{drop}$, we need an estimate for the free energy change of moving lidpids and precursors from bulk into the droplet.
For surfactants, this value can be calculated from their partition coefficient---\emph{i.e.} the fraction of lipids found in bulk solution as opposed to the aggregates.
Bachman et al. estimate this value to be 14\% for surfactants with comparable solubility \cite{Bachmann:1991} 
\[
\Delta\mu_\text{L} = k_\text{B}T \log(K_\text{L}) = k_\text{B}T \log(0.14^{-1}-1).
\]
At $T=300 \text{K}$, this evaluates to $\Delta\mu_\text{L}=7.52\times 10^{-21} \text{J}$ or $4.53 \text{kJ/mol}$.
Since the anhydride has two hydrophobic chains, we set $\Delta\mu_\text{P}=2 \Delta\mu_\text{L} = 9.06 \text{kJ/mol}$, which in turn evaluates to a partition coefficient of 2.5\%.

\begin{figure}
	\includegraphics[width=\columnwidth]{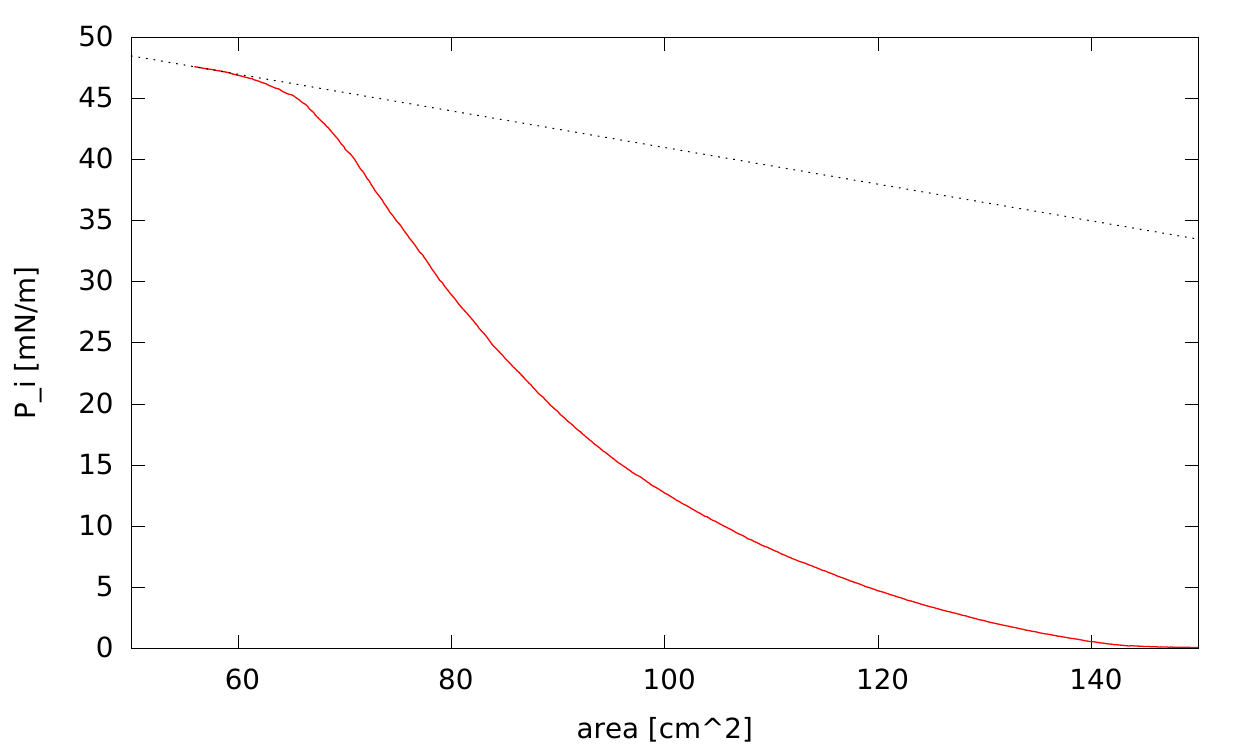}
	\caption{
	Langmuir trough measurements of a $0.5 \text{mg/ml}$ decanoic acid/POPC mixture
	with 5 mass percent decanoic acid.
	The solid red line indicates the change in pressure upon compression.
	Maximal compression is reached at $55.14 \text{cm}^2$ with $P=45.9 \text{mN/m}$.
	At this point, the infinitesimal pressure change $\partial P/\partial a$ (dotted black line) is $0.147 \text{mNm$^{-1}$/cm}^2$.
	}
	\label{fig_langmuir}
\end{figure}

The parameters $\beta$ and $\gamma$ can be deduced from Langmuir trough measurements---see figure~\ref{fig_langmuir}.
The surface tension parameter $\gamma$ is the measured line pressure at maximal compression of the lipid monolayer and equals $45.9 \text{mN/m}$.
The compressibility parameter $\beta$, determined in equation~\eqref{eq:beta}, relates to the slope $\partial P/\partial a$ at maximal compression.
In our measurements, $\partial P/\partial a$ resolves to $0.147 \text{mNm$^{-1}$/cm$^2$}$, leading to $\beta=5.80 \times 10^{-45} \text{Nm}^3$.

To fix $G_\text{system}$, we need to estimate configurational and mixing entropy components that are not yet part of the above free energies.
The mixing entropy depends on the available system volume per oil droplet as
\[
	S_\text{mixing} = n k_\text{B} \log\frac{V_\text{system}}{V_\text P/e} .
\]
To estimate the system volume, we realize that emulsion droplets of typically $100 \text{nm}$ radius have a volume of $0.0040$ femtoliter, which---assuming a water-to-oil ratio of 10:1---gives a system volume of $0.044$ femtoliter per droplet.
This also implies that a milliliter of emulsion has an order of magnitude of $10^{13}$ oil droplets.

Finally, to calculate the configurational entropy,
\[
	S_\text{config} = k_\text B \log\left[\left(\begin{matrix} L_\text{tot} \\ L\end{matrix}\right)\left(\begin{matrix} P_\text{tot} \\ P\end{matrix}\right)\right],
\]
we conclude from the ratio of precursor to droplet volume, that the droplet contains some $7,430,000$ molecules plus 2.5\% in bulk, totalling to about $P_\text{c}=7,620,000$ anhydride molecules.
With $L^\star(P_\text{d})$ and a partition coefficient of 14\%, this implies a total of $L_\text{c}=570,000$ surfactant molecules.
Expressed in concentrations, our emulsion consists of $287 \text{mmol/l}$ decanoic anhydride and $21 \text{mmol/l}$ decanoic acid.
With these numbers, we have all information required to explicitly compute the free energy $G_\text{system}$ of a certain volume of emulsion.

The numbers obtained by this calibration procedure consider the system in its relaxed state, where we have perfect covering of the droplet surface and perfect partitioning between bulk and aggregate.
When constructing the non-equilibrium thermodynamic cycle, feeding and metabolic turnover will necessarily drive the system out of this relaxed state and create tension due to compression or dilusion of the surfactant layer.
In order to introduce the least bias, we design the non-equilibrium thermodynamic cycle around our calibration point $(L_\text{c}, P_\text{c})$ such that the distance to the calibrated values is minimzed.
Assuming instantaneous feeding, the life cycle is characterized by three points: $(L_0,P_0)$ at the beginning of the process, $(L_1, P_1)$ just after feeding of $\Delta P_\text{feed}$ precursor molecules, $(L_2, P_2)$ after full metabolic conversion, and $(L_3, P_3)=(L_0,P_0)$ after extraction of surplus material.
In order to obtain a closed cycle, the following conditions must hold among these points:
\begin{align*}
	L_1 &= L_0 \\
	P_1 &= P_2 + \nu^{-1}(L_2-L_1) \\
	L_2 &= \frac{n+1}{n} L_0 \\
	P_2 &= \frac{n+1}{n} P_0
\end{align*}
Thus, an average amount of $\Delta P_\text{feed} = P_1-P_0  = \frac{1}{n} P_0 + \frac{1}{\nu n} L_0$ has to be supplied, of which $\Delta P = \frac{1}{\nu n} L_0$ are converted into $\Delta L = \frac{1}{n} L_0$ surfactants.
With the above parametrization, this implies that only a fraction $\phi=L_0/\left(\nu P_0+L_0\right) \approx 4.4\%$ of the supplied precursor is actually converted into lipids.
For any given $n$, these relations fix all but two parameters, and we choose the points in a way that they span a triangle whose center of mass coincides with $(L_\text{c}, P_\text{c})$.

We are interested in determining the number of stable droplets for each pair of $(L_\text{tot},P_\text{tot})$ values around this life cycle, or equivalently, within the $L,P$ plane.
To this end, we sample the phase space along 25 lines of constant $P_\text{tot}$ ranging from below $P_0$ to above $P_1$.
Along each line, we determine the critical amount of $L_\text{tot}$ needed, such that the minimal free energies $\left.G(n)\right|_{L_\text{tot},P_\text{tot}}$ and $\left.G(n+1)\right|_{L_\text{tot},P_\text{tot}}$ are equal for a given $n$.
This is done successively for growing $n$ until the critical points fall outside the area of observation.
By connecting points of equal $n$ between different $P_\text{tot}$ values, we obtain the borders between areas of stability for certain droplet numbers (shown as dashed lines in Figure~\ref{fig_cycle_numerics} with areas labelled with the amount of stable droplets).
When the metabolic process crosses a stability border from left to right, this indicates that the surface compression of the aggregates is too strong to support the current number of droplets and the system responds by rearranging into a configuration with one additional droplet.
When stability borders are crossed from bottom to top during the feeding process, the surplus oil phase dilutes droplet surfaces which in turn fuse into fewer aggregates.
The crucial observation is that the metabolic turnover generates one aggregate more compared to the state before feeding, and alternatively, that the step from $(L_2,P_2)$ back to $(L_0,P_0)$ crosses one stability line---which corresponds to one droplet being expelled due to the boundary conditions.
Figure~\ref{fig_cycle_numerics} shows an example phase space where $n$ has been set to 10, but we have performed calculations with $n$ ranging up to 100,000.
Calculations for other system sizes show essentially the same behavior, which is expected as the free energy is an extensive quantity.

When feeding in smaller batches or continuously instead of instantaneously, the system will traverse the state space on a path that lies within the triangle spanned by $(L_0,P_0)$, $(L_1,P_1)$ and $(L_2,P_2)$.
In the extreme case, where precursors are supplied at the same rate as they are metabolized, the metabolic path $(L_0,P_0)\rightarrow(L_2,P_2)$ coincides with the extraction path $(L_2,P_2)\rightarrow(L_0,P_0)$.
As can be seen in Figure~\ref{fig_cycle_numerics}, instantaneous feeding along the $(L_0,P_0)\rightarrow(L_1,P_1)$ path can induce droplet coalescence due to the sudden increase in lipid precursors.
Continuous feeding, on the other hand, can prevent droplet fusion and leads the system through an uninterrupted process of droplet division and extraction.
This can be important in applications where droplets are decorated with surface molecules that should not mix among different droplets, such as inheritable carriers of genetic information~\cite{Fellermann:2007b,DeClue:2009,Maurer:2011}

\begin{figure*}
	\includegraphics[width=.45\textwidth]{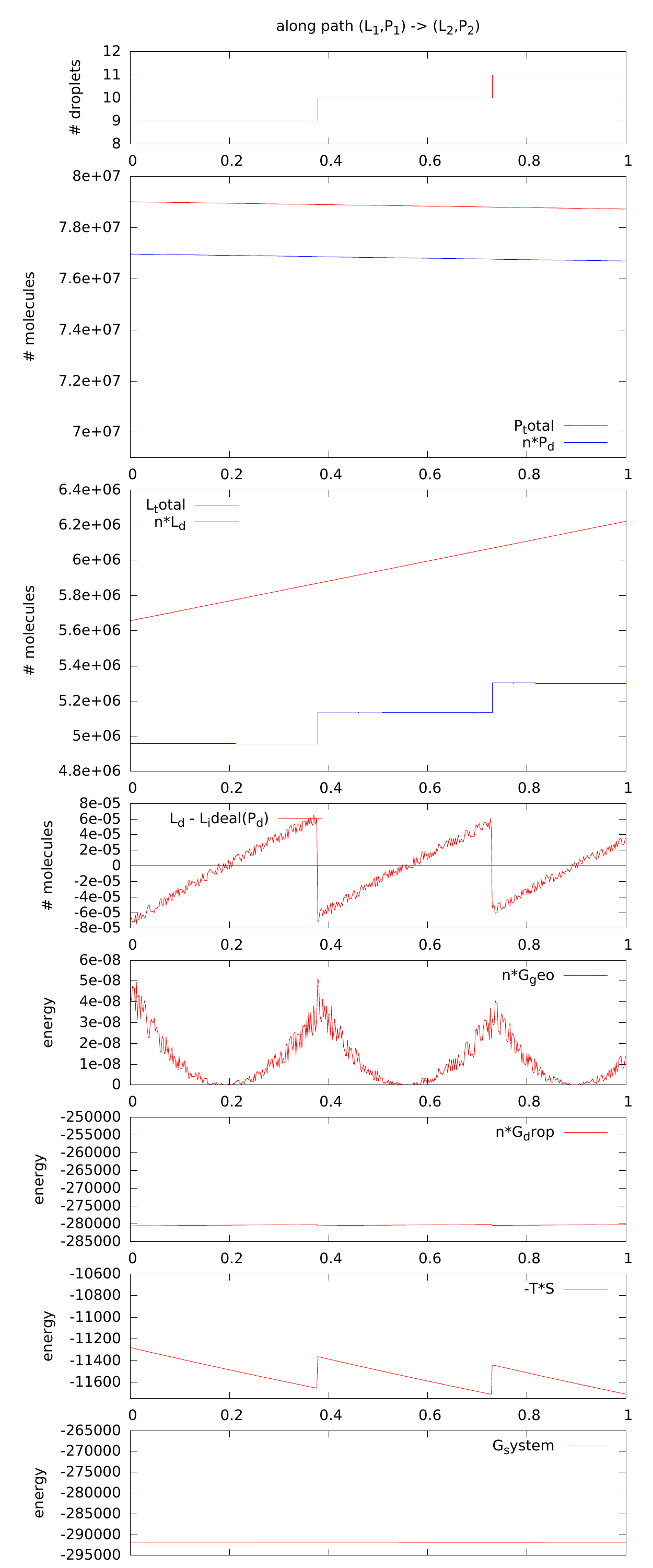}
	\includegraphics[width=.45\textwidth]{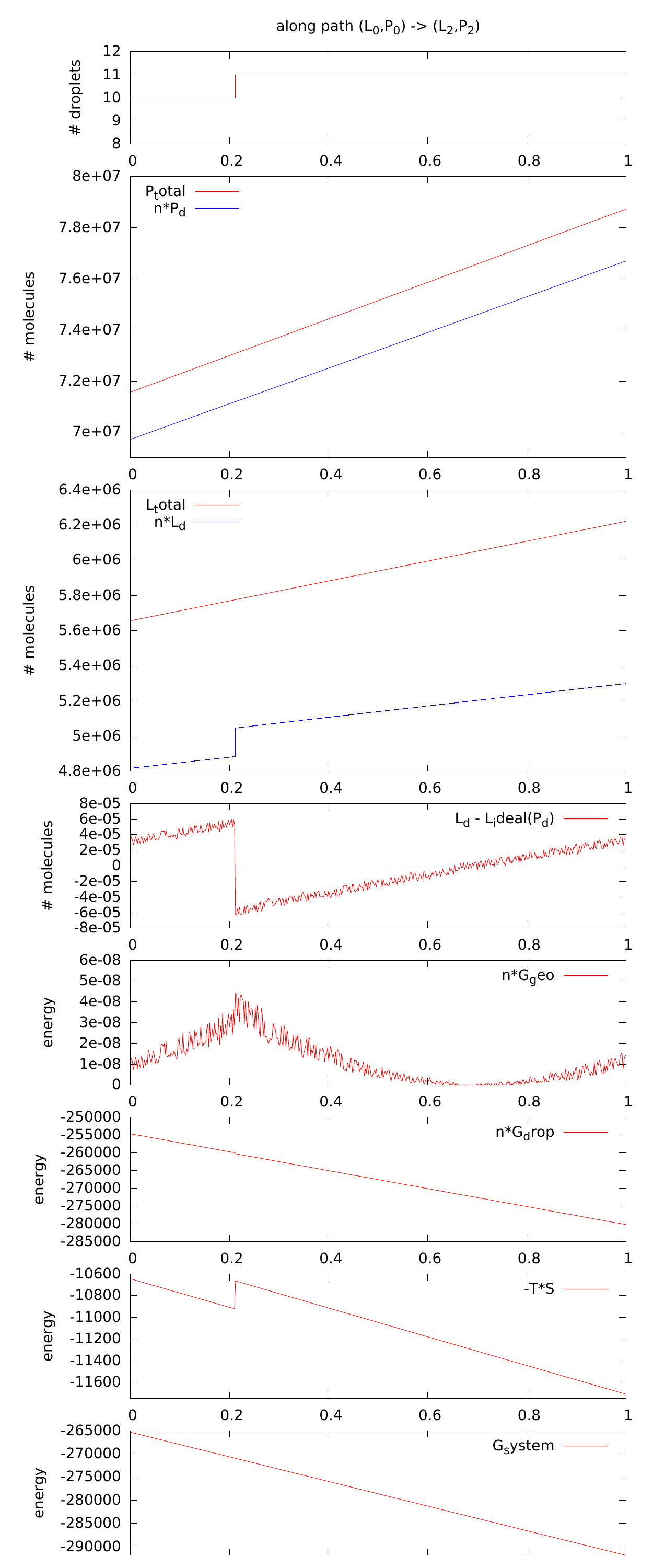}
	\caption{System state along the metabolic path assuming either instantaneous feeding (left) or continuous feeding (right).}
	\label{fig_along_path}
\end{figure*}

Figure~\ref{fig_along_path} compares these two feeding strategies in detail:
each panel shows the course of some quantity over the metabolic reaction path; $(L_1,P_1)\rightarrow(L_2,P_2)$ on the left side, and $(L_0,P_0)\rightarrow(L_2,P_2)$ on the right side.
The individual panels show:
(a) total number $n$ of droplets for a system calibrated to ten droplets at $(L_\text c, P_\text c)$,
(b) number of precursor molecules in $P_\text{tot}$ total and $nP_\text d$ in droplets,
(c) number of lipid molecules $L_\text{tot}$ in total and $nL_\text d$ in droplets,
(d) the deviation $L_\text{d} - L^\star(P_\text{d})$ from ideal surface covering of the droplets,
(e) the geometric free energy contribution,
(f) the droplet free energy contribution,
(g) the total entropy of the system,
(h) the total free energy of the system.
The trajectories emphasize that continuous feeding prevents fusion of aggregates present in the case of instantaneous feeding. 
Moreover, the results show that the surface compressibility is so strong in relation to other factors, that deviations from the ideal covering do practically not occur, and surplus lipids are instead found in bulk.
During the metabolic turnover, chemical energy is mainly used to increase the system entropy.
\end{document}